\documentclass[leter,twocolumn,10pt]{article}
\usepackage{tikz} 
\usepackage{tikz-3dplot}
\usepackage[spanish]{babel}
\usepackage[all]{xy}
\usepackage{ae}
\usepackage{graphicx}
\usepackage{amsmath}
\usepackage{anysize}  
\usepackage{amsmath}
\usepackage[latin1]{inputenc} 
\usepackage[T1]{fontenc} 
\hyphenrules{nohyphenation} 
\usepackage[]{cite}   
\usepackage{anysize}  
\marginsize{1.5cm}{1.0cm}{1.5cm}{2cm}

\begin{document}
\renewcommand{\tablename}{Tabla}
\renewcommand{\abstractname}{}
\renewcommand{\thefootnote}{\arabic{footnote}}

\title{
        {\LARGE Estimación del paso cenital solar sobre  la superficie terrestre mediante un  movimiento armónico simple}
}
\author
{ 
Paco Talero $^{1,3}$, Fernanda Santana $^{2}$,César Mora$^{3}$,\\
$^{1}$ {\small Grupo F\'{\i}sica y Matem\'{a}tica, Dpt de Ciencias
Naturales, Universidad Central},\\ 
\small {Carrera 5 No 21-38, Bogot\'{a}, D.C. Colombia.},\\ 
$^{2}$ {\small Observatorio Astronómico Nacional de Colombia}\\
\small {Universidad Nacional De Colombia,Carrera 45 No 26-85, Bogotá D.C. Colombia}\\
$^{3}$ {\small Centro de Investigaci\'{o}n en Ciencia Aplicada y Tecnolog\'{\i}a Avanzada del Instituto Polit\'{e}cnico Nacional,} \\
\small {Av. Legaria 694, Col. Irrigaci\'{o}n, C. P. 11500, M\'{e}xico D. F.}
}
\date{}

\twocolumn
[
\begin{@twocolumnfalse}

\maketitle 
\begin{abstract}
Se muestra la cinemática del paso cenital Solar sobre la Tierra como un movimiento armónico simple. Tal modelo analítico hace uso del movimiento circular uniforme para describir la rotación de la Tierra y su traslación alrededor del Sol; considera los rayos de luz Solar paralelos y la oblicuidad de la eclíptica  constante. Se confrontaron algunos resultados arrojados por el modelo analítico con los datos obtenidos del programa Stellarium 0.12.1 obteniendo acuerdos con precisión $\Delta t=1d$, lo  que permitió  proyectar el modelo  como un contexto de astronomía propicio para desarrollar la cinemática del  movimiento armónico simple.\\

\textit{Descriptores:} paso cenital, movimiento armónico simple.\\ 

The kinematics of the Solar zenith passage on the Earth was shown as a simple harmonic motion. This analytical model used the uniform circular motion to describe the Earth's rotation relative to its axis and relative to the Sun, also this model considered the sunlight beams parallel and constant the obliquity of the ecliptic. We made a contrast between some data taken from of the software Stellarium 0.12.1 and the model, we obtained an uncertainty $\Delta t=1d$; this result allow to project  the model as an astronomical context conducive to develop the kinematics of simple harmonic motion.  \\



\end{abstract}
\end{@twocolumnfalse}
]
\section{Introducción}
De acuerdo con  investigaciones recientes sobre  enseñanza de la física los contenidos físicos enmarcados en contextos astronómicos generan gran  motivación en los estudiantes que cursan  diferentes asignaturas de física en diversos estadios de formación, situación por la cual se han presentado trabajos que buscan desarrollar algunos contenidos físicos a través de distintos contextos  astronómicos\cite{PacTal}. Tales contextos contemplan tópicos como la estimación del tiempo de iluminación Solar sobre la superficie terrestre, la escala del sistema Solar, las curvas de rotación en galaxias espirales, la estimación de distancias a los planetas y  estrellas y el estudio de las leyes de Hubble y Wien \cite{Edi,Planet,Trayec,SimuF,EXV,Santa,CurvaR,Exec,Luna,Hubble,Wien}.\\
 
En la Ref.\cite{PacTal} se formula un modelo analítico que permite  estimar el tiempo de iluminación Solar sobre la Tierra para cualquier fecha del año y cualquier latitud, el modelo muestra el movimiento circular uniforme en un contexto astronómico; en la Ref.\cite{Planet} se reporta la experiencia de aula que tiene como centro de atención la búsqueda de planetas extrasolares mediante la captura y análisis de datos a través   de un telescopio dirigido a  control remoto; en la Ref.\cite{Trayec} se muestran y aclaran algunas ideas erróneas manifestadas por profesores a cerca de la trayectoria de los planetas; en las Refs.\cite{SimuF,EXV} se propone una alternativa que busca desarrollar la intuición física sobre el movimiento de cuerpos bajo interacción gravitacional y la tercera ley de Kepler a través de experimentos virtuales de fácil implementación; en la Ref.\cite{CurvaR} se proponen técnicas  para estudiar las curvas de rotación en galaxias espirales; en la Ref.\cite{Exec} se obtiene la excentricidad de la órbita terrestre usando un instrumento de observación cuyo análisis se basa en las leyes de Kepler; en Ref.\cite{Luna} mediante métodos geométricos se estiman los diámetros del Sol y la Luna así como las distancias  Tierra -Luna y Tierra -Sol; en la Ref.\cite{Hubble}  se aborda el tema de la interpretación gráfica de  velocidad contra distancia aprovechando el contexto de la ley de Hubble  y en la Ref.\cite{Wien} se muestra como la estadística de fotones al estudiar la ley de radiación de cuerpo negro es más clara y eficaz, desde el punto de vista pedagógico, que el tratamiento corriente.\\  

El problema que se aborda en este trabajo consiste en responder la pregunta: ¿cómo puede estimarse el paso cenital del Sol sobre la superficie terrestre  mediante un modelo analítico basado en la cinemática del movimiento armónico simple (MAS)?  Para esto  se desarrolla un modelo analítico cuya aproximación principal consiste en dejar de lado algunos movimientos celestes que no son relevantes durante el transcurso de unos pocos años. Así, en la Sec. ($2$) se deduce  una expresión entre la latitud y el tiempo en el cual el Sol se encuentra en el cenit, y luego se realiza la aproximación de MAS; en la Sec. ($3$) se comparan los resultados del modelo analítico  con  datos obtenidos del programa Ste\-lla\-ri\-um; en la Sec. ($4$) se calcula la precisión del modelo y se exponen algunas sugerencias pedagógicas y en la Sec. ($5$) se presentan las conclusiones.

\section{Paso cenital del Sol como un MAS}

Debido a la rotación de la Tierra sobre su propio eje esta  no tiene una geometría por completo esférica sino más bien de elipsoide, gracias a esto y a la interacción gravitacional con la Luna, el Sol y los demás cuerpos celestes se presentan los efectos de precesión  y  nutación \cite{Karttunen,Green,Portilla}. Sin embargo, estos movimientos no son apreciables durante el transcurso de unos pocos años y por tanto no son tomados en cuenta. Asimismo, debido a la baja excentricidad de la órbita terrestre, esta se considerará como una circunferencia. De otro lado, la luz Solar se toma con  la aproximación de rayos paralelos, debido a  que  observaciones bien establecidas  muestran una desviación del paralelismo de apenas $0,5^{o}$  que tiene efectos despreciables cuando se trata de proyectar sombras sobre la Tierra  \cite{Paul}.\\  

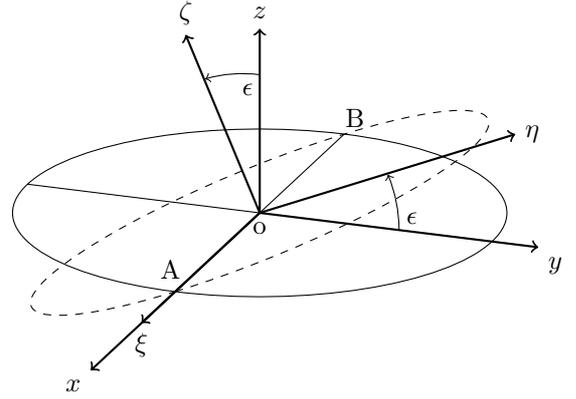
\begin{figure}[!htp]
\begin{center}
\tdplotsetmaincoords{70}{110}
\begin{tikzpicture}[scale=6.5,tdplot_main_coords]
\draw[thick,->] (0,0,0) -- (1.0,0,0) node[anchor=north east]{$x$};
\draw[thick,->] (0,0,0) -- (0,0.6,0) node[anchor=north west]{$y$};
\draw[thick,->] (0,0,0) -- (0,0,0.4) node[anchor=south]{$z$};
\draw[color=black] (-0.5,0,0)--(0,0,0);

\coordinate [label=below:\textcolor{black} {A}] (x) at  (0.28,-0.09);
\coordinate [label=below:\textcolor{black} {B}] (x) at  (-0.7,-0.05);
\coordinate [label=below:\textcolor{black} {o}] (x) at  (0.0,0.0);
\draw[color=black] (0,-0.5,0)--(0,0,0);
\tdplotdrawarc[color=black]{(0,0,0)}{0.5}{0}{360}{anchor=north west,color=black,line width=1.0pt}{}
\tdplotsetrotatedcoords{270}{23.5}{90} 
\draw[thick,color=black,tdplot_rotated_coords,->] (0,0,0) --(0.7,0,0) node[anchor=north]{$\xi$};
\draw[thick,color=black,tdplot_rotated_coords,->] (0,0,0) --(0,0.6,0) node[anchor=west]{$\eta$};
\draw[thick,color=black,tdplot_rotated_coords,->] (0,0,0) --(0,0,0.4) node[anchor=south]{$\zeta$};
\tdplotdrawarc[dashed,tdplot_rotated_coords,color=black]{(0,0,0)}{0.5}{0}{360}{anchor=north west,color=black,line width=1.0pt}{}
\tdplotsetrotatedcoords{180}{90}{270} 
\tdplotdrawarc[tdplot_rotated_coords,color=black,->]{(0,0,0)}{0.3}{270}{247}{anchor=north west,color=black,line width=1.0pt}{$\epsilon$}
\tdplotdrawarc[tdplot_rotated_coords,color=black,->]{(0,0,0)}{0.3}{0}{-23}{anchor=north west,color=black,line width=1.0pt}{$\epsilon$}

\end{tikzpicture}
\caption{Sistemas de referencia inercial}
\label{Sist}
\end{center}
\end{figure}

El sistema de referencia inercial $\left(x,y,z\right)$ escogido  tiene como origen el centro de la Tierra y se define de manera que el plano $xy$ coincida con el plano ecuatorial de la Tierra y por tanto con el plano ecuatorial celeste, el eje $z$ se elige de manera que sea paralelo al eje de rotación de la Tierra con dirección al polo norte geográfico de la Tierra y en consecuencia con dirección al polo norte celeste, además el eje $x$ se toma en dirección del punto vernal, ver Fig. \ref{Sist}.\\  

Dado que desde el sistema $\left(x,y,z\right)$  no se observa el movimiento del Sol con trayectoria circular sobre el ecuador celeste, se define otro sistema inercial denotado como  $\left(\xi,\eta,\zeta \right)$  también ligado al centro de la Tierra, que está constituido por el plano de la eclíptica rotado un ángulo $\epsilon$  respecto al ecuador celeste y sobre el cual el Sol si tiene una trayectoria circular.  En el caso de la aproximación de trayectoria circular del Sol los sistemas comparten la coordenada $x$ y $\xi$ que contienen los puntos $A$ y $B$ los cuales corresponden a los nodos ascendente y descendente respectivamente, ver Fig. \ref{Sist}.\\ 

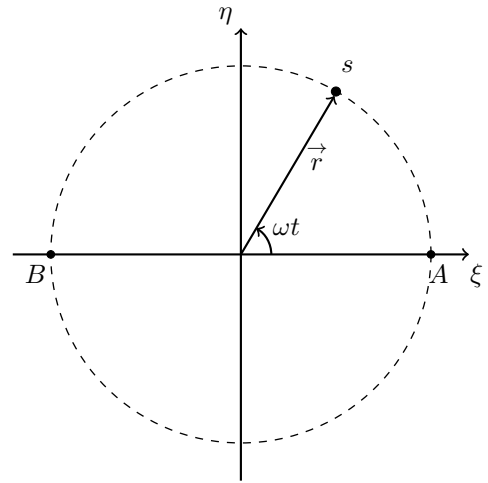
\begin{figure}[!htp]
\begin{center}
	\begin{tikzpicture}[scale=1.0]
	\draw[black,line width=0.8pt,->] (-3cm,0cm)--(3cm,0cm);
	\draw[black,line width=0.8pt,->] (0cm,-3cm)--(0cm,3cm);
	\draw [black,line width=0.5pt,dashed]   (0,0)   circle (2.5cm);
	\coordinate [label=below:\textcolor{black} {$\xi$}] (x) at  (3.1cm,0.0cm);
	\coordinate [label=below:\textcolor{black} {$\eta$}] (x) at  (-0.2cm,3.4cm);
	\coordinate [label=below:\textcolor{black} {$A$}] (x) at  (2.6cm,-0.01cm);
	\coordinate [label=below:\textcolor{black} {$B$}] (x) at  (-2.7cm,-0.01cm);
   \coordinate [label=below:\textcolor{black} {$\stackrel{\rightarrow}{r}$}] (x) at     
    (1.0cm,1.6cm);
   \coordinate [label=below:\textcolor{black} {$\omega t$}] (x) at  (0.6cm,0.6cm);
   \coordinate [label=below:\textcolor{black} {$s$}] (x) at  (1.4cm,2.7cm);
	 \draw [black, fill] (-2.5cm,0) circle (0.05cm);
	 \draw [black, fill] (2.5cm,0)  circle (0.05cm);
	 \draw [black, fill] (1.25cm,2.16cm) circle (0.06cm);
	 \draw[black,line width=0.8pt,->] (0cm,0cm)--(1.23cm,2.10cm);
	 \draw[line width=0.8pt,black,->] (0.4cm,0.0cm) arc (0:60:0.4cm);
	 
	\end{tikzpicture}
\caption{El Sol en un MCU sobre el plano de la eclíptica.}
\label{Eclip}
\end{center}
\end{figure}

En el sistema de referencia $\left(\xi,\eta,\zeta \right)$ el Sol mantiene un movimiento circular uniforme (MCU) en el plano de la eclíptica y  va del nodo  ascendente $A$ al nodo descendente  $B$ si\-gui\-en\-do la regla de la mano derecha, como muestra la Fig. \ref{Eclip}.\\ 
Ya que  el Sol tiene un MUC su posición  $\stackrel{\rightarrow}{r}\left(t\right)$ está dada por   
\begin{equation}\label{pos}
\stackrel{\rightarrow}{r}\left(t\right)=R \cos\left(\omega t\right)\widehat{u}_{\xi}+R \sin\left(\omega t\right)\widehat{u}_{\eta},
\end{equation}
donde $\widehat{u}_{\xi}$ y $\widehat{u}_{\eta}$ son vectores unitarios en dirección de los ejes $\xi$ y $\eta$ respectivamente.\\ 
  
Se quiere ahora describir el movimiento del Sol desde el sistema $\left(x,y,z\right)$, para esto es necesario relacionar los vectores $\widehat{u}_{\xi}$ y  $\widehat{u}_{\eta}$ con los vectores unitarios $\widehat{u}_{x}$,$\widehat{u}_{y}$ y $\widehat{u}_{z}$. Nótese que de acuerdo con la Fig. \ref{Sist}  
\begin{equation}\label{xi}
\widehat{u}_{\xi}=\widehat{u}_{x}
\end{equation}
y 
\begin{equation}\label{yzeta}
\widehat{u}_{\eta}=\cos(\epsilon)\widehat{u}_{y}+\sin(\epsilon)\widehat{u}_{z}.
\end{equation}
Al reemplazar las Eqs. (\ref{xi})  y (\ref{yzeta})  en la Eq. (\ref{pos})  se encuentra 
\begin{equation}\label{posE}
\begin{aligned}
\stackrel{\rightarrow}{r}\left(t\right)=R\cos\left(\omega t\right)\widehat{u}_{x}+R\sin\left(\omega t\right)\cos(\epsilon)\widehat{u}_{y}\\
+R\sin\left(\omega t\right)\sin(\epsilon) \widehat{u}_{z},
\end{aligned}
\end{equation}
donde $R$ es la distancia Tierra-Sol, $\omega$  la frecuencia angular que es $2\pi/T$ siendo $T\approx 365,25d$ el periodo de rotación del Sol en el plano de la eclíptica y $t$  el tiempo medido en días a partir del paso del Sol por el nodo ascendente $A$.\\

\begin{figure}[!htp]
\begin{center}
\tdplotsetmaincoords{70}{110}
\begin{tikzpicture}[scale=6.5,tdplot_main_coords]
\draw[thick,color=black] (0.5,0,0)--(0.15,0,0) node[anchor=west]{$o$};
\draw[thick,->] (0,0,0) -- (1.0,0,0) node[anchor=north east]{$x$};
\draw[thick,->] (0,0,0) -- (0,0.6,0) node[anchor=north west]{$y$};
\draw[thick,->] (0,0,0) -- (0,0,0.4) node[anchor=south]{$z$};
\draw[color=black] (-0.5,0,0)--(0,0,0);
\draw[color=black] (0,-0.5,0)--(0,0,0);
\tdplotdrawarc[color=black]{(0,0,0)}{0.5}{0}{360}{anchor=north west,color=black,line width=1.0pt}{}
\tdplotsetrotatedcoords{270}{23.5}{90} 
\draw[thick,color=black,tdplot_rotated_coords,->] (0,0,0) --(0.7,0,0) node[anchor=north]{$\xi$};
\draw[thick,color=black,tdplot_rotated_coords,->] (0,0,0) --(0,0.6,0) node[anchor=west]{$\eta$};
\tdplotdrawarc[dashed,tdplot_rotated_coords,color=black]{(0,0,0)}{0.5}{0}{360}{anchor=north west,color=black,line width=1.0pt}{}
\draw[thick,color=black,tdplot_rotated_coords] (0,0,0) --(0.409,0.286,0) node[anchor=west]{$s$};
\draw[thick,color=black] (0.409,0.263,0.114)--(0.409,0.263,0) node[anchor=west]{$p$};
\draw[thick,color=black] (0.409,0.263,0)--(0,0,0) node[anchor=west]{$$};
\end{tikzpicture}
\caption{Proyección de la luz Solar desde $s$ hasta $o$. }
\label{Paso}
\end{center}
\end{figure}
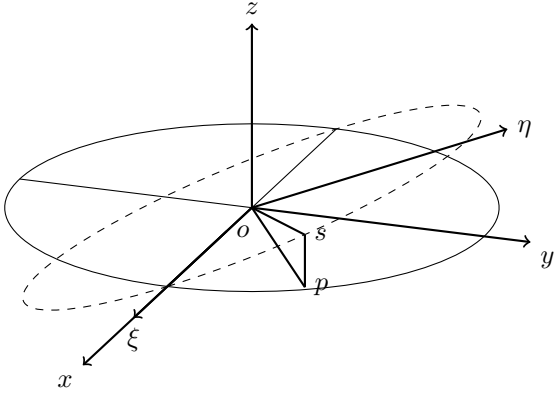

Para encontrar  la latitud $\phi$ en un  tiempo $t$   en  la cual un observador situado sobre la superficie terrestre ve el Sol en su cenit    se proyecta un rayo de luz  que viene del Sol, pasa por el cenit del observador y luego llega al centro de la Ti\-e\-rra , ver Fig. \ref{Paso}.\\ 
 
La proyección de luz sobre el centro terrestre forma el triángulo rectángulo $\widehat{spo}$ mostrado en la Fig. \ref{Paso}, el cual se ha extraído a la Fig. \ref{Tri} donde $z=\sin(\epsilon)\sin(\omega t)$. De  la geometría implícita en este triangulo se encuentra 
\begin{figure}[!htp]
\begin{center}
	\begin{tikzpicture}[scale=1.0]
	\draw[black,line width=0.8pt] (0cm,0cm)--(4cm,0cm);
	\draw[black,line width=0.8pt] (4cm,0cm)--(4cm,1.5cm);
	\draw[black,line width=0.8pt] (0cm,0cm)--(4cm,1.5cm);
	 \draw [black, fill] (0cm,0cm) circle (0.05cm);
	 \draw [black, fill] (4cm,0cm)  circle (0.05cm);
	 \draw [black, fill] (4cm,1.5cm)  circle (0.05cm);
	\coordinate [label=below:\textcolor{black} {$s$}] (x) at  (4.1cm,2.0cm);
	\coordinate [label=below:\textcolor{black} {$p$}] (x) at  (4.1cm,0.0cm);
	\coordinate [label=below:\textcolor{black} {$R$}] (x) at  (2.0cm,1.4cm);
	\coordinate [label=below:\textcolor{black} {$z$}] (x) at  (4.2cm,1.0cm);
	\coordinate [label=below:\textcolor{black} {$o$}] (x) at  (0.0cm,0.0cm);
	\coordinate [label=below:\textcolor{black} {$\phi$}] (x) at  (1.4cm,0.6cm);
	\draw[line width=0.8pt,black,->] (1.2cm,0.0cm) arc (0:20:1.2cm);	 
	\end{tikzpicture}
\caption{Proyección de la luz Solar sobre el centro terrestre.}
\label{Tri}
\end{center}
\end{figure}
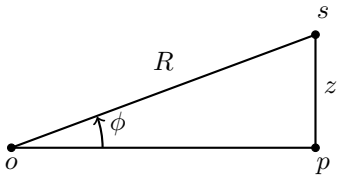
\begin{equation}\label{arc}
\sin(\phi)=\sin(\epsilon)\sin(\omega t).
\end{equation}
Tomando la aproximación de primer orden en serie de Taylor $\sin(\phi)\approx \phi$ y $\sin(\epsilon)\approx \epsilon $ la Eq.\ref{arc} se transforma en
\begin{equation}\label{mas}
\phi=\epsilon\sin(\omega t)
\end{equation}
que describe un MAS.     

\section{Precisión del modelo}

Las Eqs. (\ref{arc}) y (\ref{mas})  coinciden para tiempos  $0,T/4$,$T/2$, $3T/4$ y $T$ de manera que la diferencia entre estas formulaciones oscila entre estos tiempos. Al comparar estas ecuaciones se encuentra un error porcentual de a lo sumo $2,9\%$ que trae como consecuencia  una discrepancia máxima en la latitud de $\approx 15"$  que se presenta alrededor de los días   $36$, $146$, $219$, $329$  contados a partir del $21$ de marzo, estos días corresponden aproximadamente al $26$ abril, $13$ de agosto, $25$ de octubre y $12$ de enero respectivamente. Ahora, a  partir de la Eq.(\ref{arc}) se encuentra que  la rapidez máxima  del movimiento del Sol por el cielo, es decir $\frac{d\phi}{dt}$, es $\approx23"/d$ lo que implica como máximo una incertidumbre $\Delta t= 1d$ en la aproximación sugerida por la Eq.(\ref{mas}).\\      

\begin {figure}
\begin{center}
\setlength{\unitlength}{0.240900pt}
\ifx\plotpoint\undefined\newsavebox{\plotpoint}\fi
\sbox{\plotpoint}{\rule[-0.200pt]{0.400pt}{0.400pt}}%
\begin{picture}(1062,826)(0,0)
\sbox{\plotpoint}{\rule[-0.200pt]{0.400pt}{0.400pt}}%
\put(211.0,131.0){\rule[-0.200pt]{4.818pt}{0.400pt}}
\put(191,131){\makebox(0,0)[r]{-0.42}}
\put(991.0,131.0){\rule[-0.200pt]{4.818pt}{0.400pt}}
\put(211.0,349.0){\rule[-0.200pt]{4.818pt}{0.400pt}}
\put(191,349){\makebox(0,0)[r]{-0.14}}
\put(991.0,349.0){\rule[-0.200pt]{4.818pt}{0.400pt}}
\put(211.0,567.0){\rule[-0.200pt]{4.818pt}{0.400pt}}
\put(191,567){\makebox(0,0)[r]{ 0.14}}
\put(991.0,567.0){\rule[-0.200pt]{4.818pt}{0.400pt}}
\put(211.0,785.0){\rule[-0.200pt]{4.818pt}{0.400pt}}
\put(191,785){\makebox(0,0)[r]{ 0.42}}
\put(991.0,785.0){\rule[-0.200pt]{4.818pt}{0.400pt}}
\put(211.0,131.0){\rule[-0.200pt]{0.400pt}{4.818pt}}
\put(211,90){\makebox(0,0){$0$}}
\put(211.0,765.0){\rule[-0.200pt]{0.400pt}{4.818pt}}
\put(371.0,131.0){\rule[-0.200pt]{0.400pt}{4.818pt}}
\put(371,90){\makebox(0,0){$73$}}
\put(371.0,765.0){\rule[-0.200pt]{0.400pt}{4.818pt}}
\put(531.0,131.0){\rule[-0.200pt]{0.400pt}{4.818pt}}
\put(531,90){\makebox(0,0){$146$}}
\put(531.0,765.0){\rule[-0.200pt]{0.400pt}{4.818pt}}
\put(691.0,131.0){\rule[-0.200pt]{0.400pt}{4.818pt}}
\put(691,90){\makebox(0,0){$219$}}
\put(691.0,765.0){\rule[-0.200pt]{0.400pt}{4.818pt}}
\put(851.0,131.0){\rule[-0.200pt]{0.400pt}{4.818pt}}
\put(851,90){\makebox(0,0){$292$}}
\put(851.0,765.0){\rule[-0.200pt]{0.400pt}{4.818pt}}
\put(1011.0,131.0){\rule[-0.200pt]{0.400pt}{4.818pt}}
\put(1011,90){\makebox(0,0){$365$}}
\put(1011.0,765.0){\rule[-0.200pt]{0.400pt}{4.818pt}}
\put(211.0,131.0){\rule[-0.200pt]{0.400pt}{157.549pt}}
\put(211.0,131.0){\rule[-0.200pt]{192.720pt}{0.400pt}}
\put(1011.0,131.0){\rule[-0.200pt]{0.400pt}{157.549pt}}
\put(211.0,785.0){\rule[-0.200pt]{192.720pt}{0.400pt}}
\put(50,458){\makebox(0,0){$\rotatebox{90} { \parbox{4cm}{\begin{align*}  \phi(rad)  \end{align*}}}$}}
\put(611,29){\makebox(0,0){$t(d)$}}
\put(772,672){\makebox(0,0)[r]{$MAS$}}
\put(792.0,672.0){\rule[-0.200pt]{24.090pt}{0.400pt}}
\put(211,458){\usebox{\plotpoint}}
\multiput(211.59,458.00)(0.488,1.286){13}{\rule{0.117pt}{1.100pt}}
\multiput(210.17,458.00)(8.000,17.717){2}{\rule{0.400pt}{0.550pt}}
\multiput(219.59,478.00)(0.488,1.286){13}{\rule{0.117pt}{1.100pt}}
\multiput(218.17,478.00)(8.000,17.717){2}{\rule{0.400pt}{0.550pt}}
\multiput(227.59,498.00)(0.488,1.286){13}{\rule{0.117pt}{1.100pt}}
\multiput(226.17,498.00)(8.000,17.717){2}{\rule{0.400pt}{0.550pt}}
\multiput(235.59,518.00)(0.488,1.286){13}{\rule{0.117pt}{1.100pt}}
\multiput(234.17,518.00)(8.000,17.717){2}{\rule{0.400pt}{0.550pt}}
\multiput(243.59,538.00)(0.488,1.286){13}{\rule{0.117pt}{1.100pt}}
\multiput(242.17,538.00)(8.000,17.717){2}{\rule{0.400pt}{0.550pt}}
\multiput(251.59,558.00)(0.488,1.220){13}{\rule{0.117pt}{1.050pt}}
\multiput(250.17,558.00)(8.000,16.821){2}{\rule{0.400pt}{0.525pt}}
\multiput(259.59,577.00)(0.489,1.019){15}{\rule{0.118pt}{0.900pt}}
\multiput(258.17,577.00)(9.000,16.132){2}{\rule{0.400pt}{0.450pt}}
\multiput(268.59,595.00)(0.488,1.154){13}{\rule{0.117pt}{1.000pt}}
\multiput(267.17,595.00)(8.000,15.924){2}{\rule{0.400pt}{0.500pt}}
\multiput(276.59,613.00)(0.488,1.088){13}{\rule{0.117pt}{0.950pt}}
\multiput(275.17,613.00)(8.000,15.028){2}{\rule{0.400pt}{0.475pt}}
\multiput(284.59,630.00)(0.488,1.088){13}{\rule{0.117pt}{0.950pt}}
\multiput(283.17,630.00)(8.000,15.028){2}{\rule{0.400pt}{0.475pt}}
\multiput(292.59,647.00)(0.488,1.022){13}{\rule{0.117pt}{0.900pt}}
\multiput(291.17,647.00)(8.000,14.132){2}{\rule{0.400pt}{0.450pt}}
\multiput(300.59,663.00)(0.488,0.956){13}{\rule{0.117pt}{0.850pt}}
\multiput(299.17,663.00)(8.000,13.236){2}{\rule{0.400pt}{0.425pt}}
\multiput(308.59,678.00)(0.488,0.890){13}{\rule{0.117pt}{0.800pt}}
\multiput(307.17,678.00)(8.000,12.340){2}{\rule{0.400pt}{0.400pt}}
\multiput(316.59,692.00)(0.488,0.890){13}{\rule{0.117pt}{0.800pt}}
\multiput(315.17,692.00)(8.000,12.340){2}{\rule{0.400pt}{0.400pt}}
\multiput(324.59,706.00)(0.488,0.758){13}{\rule{0.117pt}{0.700pt}}
\multiput(323.17,706.00)(8.000,10.547){2}{\rule{0.400pt}{0.350pt}}
\multiput(332.59,718.00)(0.488,0.692){13}{\rule{0.117pt}{0.650pt}}
\multiput(331.17,718.00)(8.000,9.651){2}{\rule{0.400pt}{0.325pt}}
\multiput(340.59,729.00)(0.488,0.626){13}{\rule{0.117pt}{0.600pt}}
\multiput(339.17,729.00)(8.000,8.755){2}{\rule{0.400pt}{0.300pt}}
\multiput(348.59,739.00)(0.488,0.560){13}{\rule{0.117pt}{0.550pt}}
\multiput(347.17,739.00)(8.000,7.858){2}{\rule{0.400pt}{0.275pt}}
\multiput(356.00,748.59)(0.560,0.488){13}{\rule{0.550pt}{0.117pt}}
\multiput(356.00,747.17)(7.858,8.000){2}{\rule{0.275pt}{0.400pt}}
\multiput(365.00,756.59)(0.569,0.485){11}{\rule{0.557pt}{0.117pt}}
\multiput(365.00,755.17)(6.844,7.000){2}{\rule{0.279pt}{0.400pt}}
\multiput(373.00,763.59)(0.821,0.477){7}{\rule{0.740pt}{0.115pt}}
\multiput(373.00,762.17)(6.464,5.000){2}{\rule{0.370pt}{0.400pt}}
\multiput(381.00,768.60)(1.066,0.468){5}{\rule{0.900pt}{0.113pt}}
\multiput(381.00,767.17)(6.132,4.000){2}{\rule{0.450pt}{0.400pt}}
\multiput(389.00,772.61)(1.579,0.447){3}{\rule{1.167pt}{0.108pt}}
\multiput(389.00,771.17)(5.579,3.000){2}{\rule{0.583pt}{0.400pt}}
\put(397,775.17){\rule{1.700pt}{0.400pt}}
\multiput(397.00,774.17)(4.472,2.000){2}{\rule{0.850pt}{0.400pt}}
\put(413,775.67){\rule{1.927pt}{0.400pt}}
\multiput(413.00,776.17)(4.000,-1.000){2}{\rule{0.964pt}{0.400pt}}
\put(421,774.17){\rule{1.700pt}{0.400pt}}
\multiput(421.00,775.17)(4.472,-2.000){2}{\rule{0.850pt}{0.400pt}}
\multiput(429.00,772.95)(1.579,-0.447){3}{\rule{1.167pt}{0.108pt}}
\multiput(429.00,773.17)(5.579,-3.000){2}{\rule{0.583pt}{0.400pt}}
\multiput(437.00,769.93)(0.821,-0.477){7}{\rule{0.740pt}{0.115pt}}
\multiput(437.00,770.17)(6.464,-5.000){2}{\rule{0.370pt}{0.400pt}}
\multiput(445.00,764.93)(0.671,-0.482){9}{\rule{0.633pt}{0.116pt}}
\multiput(445.00,765.17)(6.685,-6.000){2}{\rule{0.317pt}{0.400pt}}
\multiput(453.00,758.93)(0.645,-0.485){11}{\rule{0.614pt}{0.117pt}}
\multiput(453.00,759.17)(7.725,-7.000){2}{\rule{0.307pt}{0.400pt}}
\multiput(462.59,750.72)(0.488,-0.560){13}{\rule{0.117pt}{0.550pt}}
\multiput(461.17,751.86)(8.000,-7.858){2}{\rule{0.400pt}{0.275pt}}
\multiput(470.59,741.72)(0.488,-0.560){13}{\rule{0.117pt}{0.550pt}}
\multiput(469.17,742.86)(8.000,-7.858){2}{\rule{0.400pt}{0.275pt}}
\multiput(478.59,732.30)(0.488,-0.692){13}{\rule{0.117pt}{0.650pt}}
\multiput(477.17,733.65)(8.000,-9.651){2}{\rule{0.400pt}{0.325pt}}
\multiput(486.59,721.09)(0.488,-0.758){13}{\rule{0.117pt}{0.700pt}}
\multiput(485.17,722.55)(8.000,-10.547){2}{\rule{0.400pt}{0.350pt}}
\multiput(494.59,709.09)(0.488,-0.758){13}{\rule{0.117pt}{0.700pt}}
\multiput(493.17,710.55)(8.000,-10.547){2}{\rule{0.400pt}{0.350pt}}
\multiput(502.59,696.68)(0.488,-0.890){13}{\rule{0.117pt}{0.800pt}}
\multiput(501.17,698.34)(8.000,-12.340){2}{\rule{0.400pt}{0.400pt}}
\multiput(510.59,682.47)(0.488,-0.956){13}{\rule{0.117pt}{0.850pt}}
\multiput(509.17,684.24)(8.000,-13.236){2}{\rule{0.400pt}{0.425pt}}
\multiput(518.59,667.47)(0.488,-0.956){13}{\rule{0.117pt}{0.850pt}}
\multiput(517.17,669.24)(8.000,-13.236){2}{\rule{0.400pt}{0.425pt}}
\multiput(526.59,652.26)(0.488,-1.022){13}{\rule{0.117pt}{0.900pt}}
\multiput(525.17,654.13)(8.000,-14.132){2}{\rule{0.400pt}{0.450pt}}
\multiput(534.59,636.06)(0.488,-1.088){13}{\rule{0.117pt}{0.950pt}}
\multiput(533.17,638.03)(8.000,-15.028){2}{\rule{0.400pt}{0.475pt}}
\multiput(542.59,618.85)(0.488,-1.154){13}{\rule{0.117pt}{1.000pt}}
\multiput(541.17,620.92)(8.000,-15.924){2}{\rule{0.400pt}{0.500pt}}
\multiput(550.59,600.85)(0.488,-1.154){13}{\rule{0.117pt}{1.000pt}}
\multiput(549.17,602.92)(8.000,-15.924){2}{\rule{0.400pt}{0.500pt}}
\multiput(558.59,583.08)(0.489,-1.077){15}{\rule{0.118pt}{0.944pt}}
\multiput(557.17,585.04)(9.000,-17.040){2}{\rule{0.400pt}{0.472pt}}
\multiput(567.59,563.64)(0.488,-1.220){13}{\rule{0.117pt}{1.050pt}}
\multiput(566.17,565.82)(8.000,-16.821){2}{\rule{0.400pt}{0.525pt}}
\multiput(575.59,544.43)(0.488,-1.286){13}{\rule{0.117pt}{1.100pt}}
\multiput(574.17,546.72)(8.000,-17.717){2}{\rule{0.400pt}{0.550pt}}
\multiput(583.59,524.43)(0.488,-1.286){13}{\rule{0.117pt}{1.100pt}}
\multiput(582.17,526.72)(8.000,-17.717){2}{\rule{0.400pt}{0.550pt}}
\multiput(591.59,504.43)(0.488,-1.286){13}{\rule{0.117pt}{1.100pt}}
\multiput(590.17,506.72)(8.000,-17.717){2}{\rule{0.400pt}{0.550pt}}
\multiput(599.59,484.43)(0.488,-1.286){13}{\rule{0.117pt}{1.100pt}}
\multiput(598.17,486.72)(8.000,-17.717){2}{\rule{0.400pt}{0.550pt}}
\multiput(607.59,464.43)(0.488,-1.286){13}{\rule{0.117pt}{1.100pt}}
\multiput(606.17,466.72)(8.000,-17.717){2}{\rule{0.400pt}{0.550pt}}
\multiput(615.59,444.43)(0.488,-1.286){13}{\rule{0.117pt}{1.100pt}}
\multiput(614.17,446.72)(8.000,-17.717){2}{\rule{0.400pt}{0.550pt}}
\multiput(623.59,424.23)(0.488,-1.352){13}{\rule{0.117pt}{1.150pt}}
\multiput(622.17,426.61)(8.000,-18.613){2}{\rule{0.400pt}{0.575pt}}
\multiput(631.59,403.64)(0.488,-1.220){13}{\rule{0.117pt}{1.050pt}}
\multiput(630.17,405.82)(8.000,-16.821){2}{\rule{0.400pt}{0.525pt}}
\multiput(639.59,384.43)(0.488,-1.286){13}{\rule{0.117pt}{1.100pt}}
\multiput(638.17,386.72)(8.000,-17.717){2}{\rule{0.400pt}{0.550pt}}
\multiput(647.59,364.64)(0.488,-1.220){13}{\rule{0.117pt}{1.050pt}}
\multiput(646.17,366.82)(8.000,-16.821){2}{\rule{0.400pt}{0.525pt}}
\multiput(655.59,346.08)(0.489,-1.077){15}{\rule{0.118pt}{0.944pt}}
\multiput(654.17,348.04)(9.000,-17.040){2}{\rule{0.400pt}{0.472pt}}
\multiput(664.59,326.85)(0.488,-1.154){13}{\rule{0.117pt}{1.000pt}}
\multiput(663.17,328.92)(8.000,-15.924){2}{\rule{0.400pt}{0.500pt}}
\multiput(672.59,308.85)(0.488,-1.154){13}{\rule{0.117pt}{1.000pt}}
\multiput(671.17,310.92)(8.000,-15.924){2}{\rule{0.400pt}{0.500pt}}
\multiput(680.59,291.06)(0.488,-1.088){13}{\rule{0.117pt}{0.950pt}}
\multiput(679.17,293.03)(8.000,-15.028){2}{\rule{0.400pt}{0.475pt}}
\multiput(688.59,274.06)(0.488,-1.088){13}{\rule{0.117pt}{0.950pt}}
\multiput(687.17,276.03)(8.000,-15.028){2}{\rule{0.400pt}{0.475pt}}
\multiput(696.59,257.47)(0.488,-0.956){13}{\rule{0.117pt}{0.850pt}}
\multiput(695.17,259.24)(8.000,-13.236){2}{\rule{0.400pt}{0.425pt}}
\multiput(704.59,242.47)(0.488,-0.956){13}{\rule{0.117pt}{0.850pt}}
\multiput(703.17,244.24)(8.000,-13.236){2}{\rule{0.400pt}{0.425pt}}
\multiput(712.59,227.68)(0.488,-0.890){13}{\rule{0.117pt}{0.800pt}}
\multiput(711.17,229.34)(8.000,-12.340){2}{\rule{0.400pt}{0.400pt}}
\multiput(720.59,214.09)(0.488,-0.758){13}{\rule{0.117pt}{0.700pt}}
\multiput(719.17,215.55)(8.000,-10.547){2}{\rule{0.400pt}{0.350pt}}
\multiput(728.59,202.09)(0.488,-0.758){13}{\rule{0.117pt}{0.700pt}}
\multiput(727.17,203.55)(8.000,-10.547){2}{\rule{0.400pt}{0.350pt}}
\multiput(736.59,190.30)(0.488,-0.692){13}{\rule{0.117pt}{0.650pt}}
\multiput(735.17,191.65)(8.000,-9.651){2}{\rule{0.400pt}{0.325pt}}
\multiput(744.59,179.51)(0.488,-0.626){13}{\rule{0.117pt}{0.600pt}}
\multiput(743.17,180.75)(8.000,-8.755){2}{\rule{0.400pt}{0.300pt}}
\multiput(752.00,170.93)(0.494,-0.488){13}{\rule{0.500pt}{0.117pt}}
\multiput(752.00,171.17)(6.962,-8.000){2}{\rule{0.250pt}{0.400pt}}
\multiput(760.00,162.93)(0.645,-0.485){11}{\rule{0.614pt}{0.117pt}}
\multiput(760.00,163.17)(7.725,-7.000){2}{\rule{0.307pt}{0.400pt}}
\multiput(769.00,155.93)(0.671,-0.482){9}{\rule{0.633pt}{0.116pt}}
\multiput(769.00,156.17)(6.685,-6.000){2}{\rule{0.317pt}{0.400pt}}
\multiput(777.00,149.93)(0.821,-0.477){7}{\rule{0.740pt}{0.115pt}}
\multiput(777.00,150.17)(6.464,-5.000){2}{\rule{0.370pt}{0.400pt}}
\multiput(785.00,144.94)(1.066,-0.468){5}{\rule{0.900pt}{0.113pt}}
\multiput(785.00,145.17)(6.132,-4.000){2}{\rule{0.450pt}{0.400pt}}
\put(793,140.17){\rule{1.700pt}{0.400pt}}
\multiput(793.00,141.17)(4.472,-2.000){2}{\rule{0.850pt}{0.400pt}}
\put(801,138.67){\rule{1.927pt}{0.400pt}}
\multiput(801.00,139.17)(4.000,-1.000){2}{\rule{0.964pt}{0.400pt}}
\put(405.0,777.0){\rule[-0.200pt]{1.927pt}{0.400pt}}
\put(817,139.17){\rule{1.700pt}{0.400pt}}
\multiput(817.00,138.17)(4.472,2.000){2}{\rule{0.850pt}{0.400pt}}
\put(825,141.17){\rule{1.700pt}{0.400pt}}
\multiput(825.00,140.17)(4.472,2.000){2}{\rule{0.850pt}{0.400pt}}
\multiput(833.00,143.60)(1.066,0.468){5}{\rule{0.900pt}{0.113pt}}
\multiput(833.00,142.17)(6.132,4.000){2}{\rule{0.450pt}{0.400pt}}
\multiput(841.00,147.59)(0.671,0.482){9}{\rule{0.633pt}{0.116pt}}
\multiput(841.00,146.17)(6.685,6.000){2}{\rule{0.317pt}{0.400pt}}
\multiput(849.00,153.59)(0.671,0.482){9}{\rule{0.633pt}{0.116pt}}
\multiput(849.00,152.17)(6.685,6.000){2}{\rule{0.317pt}{0.400pt}}
\multiput(857.00,159.59)(0.560,0.488){13}{\rule{0.550pt}{0.117pt}}
\multiput(857.00,158.17)(7.858,8.000){2}{\rule{0.275pt}{0.400pt}}
\multiput(866.59,167.00)(0.488,0.560){13}{\rule{0.117pt}{0.550pt}}
\multiput(865.17,167.00)(8.000,7.858){2}{\rule{0.400pt}{0.275pt}}
\multiput(874.59,176.00)(0.488,0.626){13}{\rule{0.117pt}{0.600pt}}
\multiput(873.17,176.00)(8.000,8.755){2}{\rule{0.400pt}{0.300pt}}
\multiput(882.59,186.00)(0.488,0.692){13}{\rule{0.117pt}{0.650pt}}
\multiput(881.17,186.00)(8.000,9.651){2}{\rule{0.400pt}{0.325pt}}
\multiput(890.59,197.00)(0.488,0.758){13}{\rule{0.117pt}{0.700pt}}
\multiput(889.17,197.00)(8.000,10.547){2}{\rule{0.400pt}{0.350pt}}
\multiput(898.59,209.00)(0.488,0.890){13}{\rule{0.117pt}{0.800pt}}
\multiput(897.17,209.00)(8.000,12.340){2}{\rule{0.400pt}{0.400pt}}
\multiput(906.59,223.00)(0.488,0.890){13}{\rule{0.117pt}{0.800pt}}
\multiput(905.17,223.00)(8.000,12.340){2}{\rule{0.400pt}{0.400pt}}
\multiput(914.59,237.00)(0.488,0.956){13}{\rule{0.117pt}{0.850pt}}
\multiput(913.17,237.00)(8.000,13.236){2}{\rule{0.400pt}{0.425pt}}
\multiput(922.59,252.00)(0.488,1.022){13}{\rule{0.117pt}{0.900pt}}
\multiput(921.17,252.00)(8.000,14.132){2}{\rule{0.400pt}{0.450pt}}
\multiput(930.59,268.00)(0.488,1.022){13}{\rule{0.117pt}{0.900pt}}
\multiput(929.17,268.00)(8.000,14.132){2}{\rule{0.400pt}{0.450pt}}
\multiput(938.59,284.00)(0.488,1.088){13}{\rule{0.117pt}{0.950pt}}
\multiput(937.17,284.00)(8.000,15.028){2}{\rule{0.400pt}{0.475pt}}
\multiput(946.59,301.00)(0.488,1.154){13}{\rule{0.117pt}{1.000pt}}
\multiput(945.17,301.00)(8.000,15.924){2}{\rule{0.400pt}{0.500pt}}
\multiput(954.59,319.00)(0.489,1.077){15}{\rule{0.118pt}{0.944pt}}
\multiput(953.17,319.00)(9.000,17.040){2}{\rule{0.400pt}{0.472pt}}
\multiput(963.59,338.00)(0.488,1.220){13}{\rule{0.117pt}{1.050pt}}
\multiput(962.17,338.00)(8.000,16.821){2}{\rule{0.400pt}{0.525pt}}
\multiput(971.59,357.00)(0.488,1.220){13}{\rule{0.117pt}{1.050pt}}
\multiput(970.17,357.00)(8.000,16.821){2}{\rule{0.400pt}{0.525pt}}
\multiput(979.59,376.00)(0.488,1.286){13}{\rule{0.117pt}{1.100pt}}
\multiput(978.17,376.00)(8.000,17.717){2}{\rule{0.400pt}{0.550pt}}
\multiput(987.59,396.00)(0.488,1.286){13}{\rule{0.117pt}{1.100pt}}
\multiput(986.17,396.00)(8.000,17.717){2}{\rule{0.400pt}{0.550pt}}
\multiput(995.59,416.00)(0.488,1.286){13}{\rule{0.117pt}{1.100pt}}
\multiput(994.17,416.00)(8.000,17.717){2}{\rule{0.400pt}{0.550pt}}
\multiput(1003.59,436.00)(0.488,1.286){13}{\rule{0.117pt}{1.100pt}}
\multiput(1002.17,436.00)(8.000,17.717){2}{\rule{0.400pt}{0.550pt}}
\put(809.0,139.0){\rule[-0.200pt]{1.927pt}{0.400pt}}
\put(772,631){\makebox(0,0)[r]{1}}
\put(351,738){\makebox(0,0){$\bullet$}}
\put(476,738){\makebox(0,0){$\bullet$}}
\put(842,631){\makebox(0,0){$\bullet$}}
\put(772,590){\makebox(0,0)[r]{2}}
\put(695,279){\makebox(0,0){$\circ$}}
\put(934,279){\makebox(0,0){$\circ$}}
\put(842,590){\makebox(0,0){$\circ$}}
\put(772,549){\makebox(0,0)[r]{3}}
\put(239,536){\makebox(0,0){$\triangle$}}
\put(584,536){\makebox(0,0){$\triangle$}}
\put(842,549){\makebox(0,0){$\triangle$}}
\put(211.0,131.0){\rule[-0.200pt]{0.400pt}{157.549pt}}
\put(211.0,131.0){\rule[-0.200pt]{192.720pt}{0.400pt}}
\put(1011.0,131.0){\rule[-0.200pt]{0.400pt}{157.549pt}}
\put(211.0,785.0){\rule[-0.200pt]{192.720pt}{0.400pt}}
\end{picture}

\caption{Comparación del MAS con resultados arrojados por el programa Stellarium 0.12.1 para: $1$ Chichén Itzá, $2$ Machu Picchu  y $3$ El Infiernito.}
\label{Mod}
\end{center}
\end {figure}
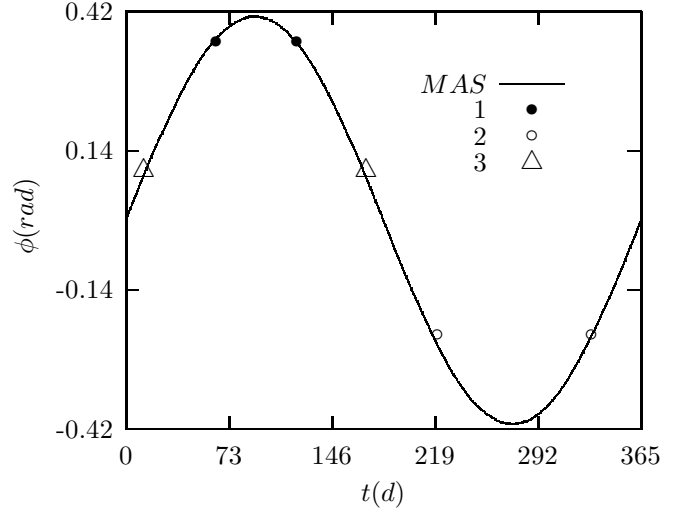

Se comparó  el modelo de MAS con datos obtenidos del programa Stellarium 0.12.1  para los años  $2013-2014$  en  algunos lugares de importancia cultural de América Latina relacionados con Arqueoastronomía, se escogieron: Chichén Itzá en Yucatán (México), Machu Picchu en cusco (Perú) y El infiernito en Villa de Leyva  (Colombia), los resultados obtenidos se muestran en la gráfica de la Fig. \ref{Mod}, la cual evidencia  acuerdo dentro de la incertidumbre $\Delta t= 1d$, encontrada anteriormente.

\section{Discusión}

De acuerdo con lo anterior, el paso cenital del Sol es un fenómeno que ocurre una vez cada año en los extremos de latitud $\phi =\pm \epsilon$ que coincide con el Solsticio de verano tanto en el hemisferio norte como  en el hemisferio sur, mientras que entre los trópicos ocurre dos veces por año y para latitudes mayores  a $\epsilon$ y menores a $\epsilon$ no ocurre. El modelo de MAS comienza en el equinoccio de primavera para el hemisferio norte alrededor del $21$ de marzo en $t=0$, cuando el Sol se mueve  rápido por él cielo; luego pasa al Solsticio de verano alrededor del $21$ de junio, cuando $t=T/4$ y el Sol se mueve por el cielo lentamente; después va al equinoccio de otoño aproximadamente el $21$ de septiembre, en un tiempo $t=T/2$; luego pasa al Solsticio de invierno hacia el $21$ de diciembre cuando $t=T/4$ y finalmente regresa al equinoccio de  primavera justo cuando $t=T$. Cuando ocurren los Solsticios la luz proveniente del Sol se encuentra únicamente en el plano de la eclíptica, mientras que cuando ocurren los equinoccios la luz se encuentra en la intercesión del plano del ecuador celeste y el de la eclíptica, es decir la luz se encuentra en la línea nodal.\\

El estudio del paso cenital del Sol considerado como un MAS permite ampliar los casos de estudio  provistos de contenido auténtico en el contenido del movimiento armónico simple,  que se imparte desde cursos de bachillerato hasta  cursos universitarios de ciencias e ingeniería. El trabajo presentado aquí permite desarrollar proyectos de aula con uno o dos semestres de duración donde se sigue el Sol por el cielo mediante la sombra de algún objeto ubicado en la superficie terrestre. Esto permite, realizar analogías con los sistemas masa-resorte, péndulo simple, entre otros. Además, facilita la implementación de proyectos  colaborativos entre escuelas de diferentes países creando comunidades científicas escolares a través de la Web. Al mismo tiempo, permite crear vínculos entre la física y  otras disciplinas como la Arqueoastronomía al mostrar una manera práctica de estimar  la fecha del año del Sol cenital.\\    

\section{Conclusiones}
Se demostró que al considerar la trayectoria de la Tierra alrededor del Sol circular, el plano de la eclíptica constante y los rayos de luz Solar paralelos entre sí, el paso cenital del Sol sobre la superficie terrestre se puede aproximar mediante un movimiento armónico simple. El modelo se confrontó con datos extraídos del programa Stellarium versión 0.12.1  obteniendo resultados acordes con una incertidumbre en el tiempo de un día, esto permitió proponer el modelo como posible contexto para el desarrollo de proyectos de aula en la clase de física ya sea de bachillerato o primeros cursos de ciencias e ingeniería.            

\section*{Agradecimientos}
Los autores agradecen a la Facultad de Ingeniería y al Departamento de Ciencias Naturales de la Universidad  Central por el tiempo y los recursos asignados al proyecto de investigación: Un modelo de enseñanza de la física mediante videos de experimentos Discrepantes realizado durante el año $2013$. Adicionalmente agradecen al observatorio astronómico  Nacional de Colombia y al CICATA del IPN de México por su continua colaboración.   
\renewcommand{\refname}{Referencias}

\end{document}